\documentclass[twocolumn]{aastex701}

\usepackage{amsmath}
\usepackage{graphicx}
\usepackage{subcaption}
\usepackage{comment}

\begin{document}


\title{Discovery of the Slowest Radio Pulsar in the Small Magellanic Cloud}

\author[orcid=0009-0007-0178-2322]{Weiyu Wang}
\affiliation{Trinity College, University of Cambridge, Cambridge CB2 1TQ, United Kingdom}
\affiliation{Max Planck Institute for Gravitational Physics (Albert Einstein Institute), D-30167 Hannover, Germany \\}
\affiliation{Leibniz Universit\"{a}t Hannover, D-30167 Hannover, Germany\\}
\email[show]{ww429@cam.ac.uk}  

\author[0000-0001-6295-2881]{Rahul Sengar} 
\affiliation{Max Planck Institute for Gravitational Physics (Albert Einstein Institute), D-30167 Hannover, Germany \\}
\affiliation{Leibniz Universit\"{a}t Hannover, D-30167 Hannover, Germany\\}
\email{rsengar@aei.mpg.de}

\begin{abstract}

We report the discovery of a new radio pulsar PSR J0103$-$72 in the Small Magellanic Cloud (SMC) through a reanalysis of Parkes/Murriyang archival observations targeting supernova remnants and pulsar-wind nebulae in the SMC. Previously two pulsars were discovered in these observations, and PSR J0103$-$72 is the third which we independently detected in two observations of its target field. It has a spin period of 1.327~s and a dispersion measure (DM) of $272\pm13~\mathrm{pc\,cm^{-3}}$, making it the slowest known radio pulsar in the SMC with the second-highest DM. We did not identify any plausible supernova remnant or pulsar-wind nebula associated with the pulsar. Subtracting the predicted Galactic foreground leaves a DM contribution of $\sim 230$--$242~\mathrm{pc\,cm^{-3}}$, which may suggest a long path through the ionized medium of the SMC, enhanced ionized gas along the sightline, or both. Although these data had previously been searched using both fast Fourier transform (FFT) and fast folding algorithms, our FFT-based reprocessing found PSR J0103$-$72 through search techniques which have previously been highly effective in discovering a large number of pulsars in other surveys. This discovery shows that candidates with weak Fourier-domain significance can reveal faint, narrow-duty-cycle pulsars missed in earlier extragalactic pulsar searches.

\end{abstract}



\section{Introduction}

The Large and Small Magellanic Clouds (LMC and SMC) are the nearest external galaxies and the only ones in which radio pulsars have been discovered. Out of the nearly 4400 previously discovered pulsars, only 66 are extragalactic, of which 63 are radio pulsars. Among these, 14 are in the SMC and 49 are in the LMC based on version 2.8.1 of the ATNF Pulsar Catalogue\footnote{\url{https://www.atnf.csiro.au/research/pulsar/psrcat}} and subsequent discoveries by \citet{prayag2024, prayag2026, sengar_lmc_2026}. The remaining three are rotation-powered pulsars detected only at X-ray wavelengths.

Despite the large distances of $\sim 50 \,\mathrm{kpc}$ for the LMC \citep{pietrzynski2019} and $\sim 62 \, \mathrm{kpc}$ for the SMC \citep{graczyk2020}, their high Galactic latitudes provide relatively clean sightlines as compared to those with directions close to the Galactic plane, with less Galactic dispersion and scattering. This enables sensitive searches for pulsars in both galaxies.

Among the Magellanic Clouds, the SMC is a particularly important target for studying neutron-star formation and evolution in a galactic environment different from that of the Milky Way. Its metallicity is substantially lower, $Z_{\mathrm{SMC}} \sim 0.2\,Z_{\mathrm{MW}}$ \citep{luck1998}, which can lead to an increased neutron star birth rate \citep{heger2003}. The SMC also contains a substantial population of high-mass X-ray binaries (HMXBs) attributed to episodes of recent star formation history \citep{zaritsky2004, mineo2012} and lower metallicity \citep{dray2006}, many of which host accreting neutron stars observed as X-ray pulsars \citep{haberl2016}. \citet{haberl2016} catalogued 148 confirmed and candidate HMXBs in the SMC, with X-ray pulsations detected from approximately half of the relatively high-confidence systems. This HMXB population demonstrates that the SMC has produced a substantial population of young neutron stars, motivating searches for its much less well-characterised radio-pulsar population.

The SMC also hosts a large population of supernova remnants (SNRs), which indicate recent supernova activity and provide targets for searches for young neutron stars, as core-collapse supernovae (CCSNe) can leave behind neutron stars or black holes \citep{heger2003}. Currently, 24 confirmed SNRs and 10 additional candidates are known in the SMC \citep{cotton2024}.The total (core-collapse and Ia) SN rate in the MCs is 1.7--3.1\,SNuM [SNe\,(100\,yr\,$10^{10}$\,M$_\odot$)$^{-1}$] \citep{maoz2010}. While type Ia supernovae do not yield neutron stars, an estimated 62--92 per cent of the SMC SNR population originated from CCSNe \citep{maggi2019}. It could be implied, as a rough estimate, that the CCSN rate in the SMC is an order of magnitude above the CCSN rate in the Milky Way of 1.63\,$\pm$\,0.46\,CCSNe per century \citep{rozwadowska2021}, which is equivalent to $\sim$0.3\,SNuM assuming a Galactic stellar mass of $(6.08 \pm 1.14) \times 10^{10}$\,M$_\odot$ \citep{licquia2015}, indicating a substantially more active star-forming environment. The large core-collapse SNR population therefore provides a set of physically motivated targets for searches for young pulsars. During the observable lifetime of an SNR, a young neutron star may remain within or near the remnant, although it can be displaced from the explosion site by its natal velocity \citep{gaensler1995,hobbs2005}. If the neutron star is sufficiently young and energetic, its rotational-energy loss can also power a pulsar wind nebula (PWN) \citep{gaensler2006}. SNRs containing PWNe are therefore particularly promising targets for targeted radio pulsar searches, since the PWN provides evidence for the presence of an active, energetic neutron star. However, only a small number of radio pulsars have been discovered in the SMC, motivating further searches for its radio-pulsar population.

Searches for radio pulsars in the SMC have progressed through almost four decades of increasing surveys sensitivity. The first radio pulsar known in the SMC\, J0045$-$7319, was reported by \citet{mcconnell1991}. The subsequent searches by \citet{crawford2001} found another pulsar after a decade, and three were later found by \citet{manchester2006}. These surveys were primarily large area searches conducted with Murriyang, the Parkes 64-m radio telescope and were sensitive only to the brighter end of the SMC pulsar luminosity distribution.

However, in recent years, deep targeted searches of SNRs, SNR candidates and PWN regions in the SMC have been successful in finding pulsars. Using Murriyang, \citet{titus2019} conducted a targeted search of such regions and found two new pulsars. With the advent of highly sensitive telescopes such as MeerKAT, the TRAPUM SMC targeted searches found 7 new pulsars which doubled the radio pulsar population in the SMC \citep{carli2024}, while boosting the sensitivity to about twice the sensitivities of the previous surveys conducted with Murriyang. This survey also found the first radio pulsar associated with PWN in the SMC. Apart from this, PSR J0040$-$7337, with a spin period of $\sim$59 ms is associated with the bow-shock PWN in the SNR DEM S5, and PSR J0048$-$7317, with a period of $\sim$79 ms is associated with another radio PWN with no known SNR association. These discoveries provide highly credible observational support for using SNRs and PWNe as high-priority targets for searches for young extragalactic pulsars.

However, increasing telescope sensitivity does not guarantee that all detectable pulsars present in a survey will be discovered. The completeness of pulsar searches also largely depends on how the data are processed. Reprocessing archival survey data with improved search algorithms, computing resources, and candidate-selection strategies has repeatedly led to the discovery of many pulsars. The reprocessing of the Parkes Multibeam Pulsar Survey is one of the prime examples in which over two hundred pulsars were found  \citep{keith2009,eatough2013,knispel2013}. More recently, \citet{sengar2023} discovered 37 additional pulsars through GPU-accelerated reprocessing of the same survey. Reanalyses of archival data have also led to the discovery of interesting systems. For example, \citet{sengar2025} reported 71 new pulsars from reprocessing approximately 87\% of the HTRU-S LowLat survey in which a double neutron star was also found \citep{sengar2022}. The importance of reprocessing have not only been limited to finding Galactic pulsars,  they have also been successful for pulsar searches in the Magellanic Clouds. \citet{ridley2013} reprocessed archival data of the Magellanic Clouds survey conducted by \citet{manchester2006} and discovered five additional radio pulsars in the LMC. Reprocessing of high-resolution Parkes observations of the LMC also discovered PSR J0556$-$67 \citep{hisano2022}. More recently, \citet{xia2025} reprocessed archival data from the Parkes 70-cm survey and discovered PSR J0540$-$69 in the LMC. These results show that archived data have significant discovery potential when reprocessed with improved search methods.

In this letter, we present the discovery of a new radio pulsar in the SMC from a reprocessing of the \citet{titus2019} targeted search data with an FFT based search pipeline optimised for discovering narrow duty cycle pulsars. In Section 2, we introduce the searches that were performed. In Section 3, we explain the data reduction stage, including parameter search ranges. Our results are then presented in Section 4 and discussed in Section 5. Finally, a brief summary is given in Section 6 to conclude this work.

\begin{deluxetable}{lcccc}
\tablecaption{Known pulsars in the SMC, ordered by spin period. Membership
  follows the ATNF Pulsar Catalogue \citep{manchester2005} classification of
  extragalactic association. Flux densities are the discovery values where
  published.\label{tab:smc}}

\tablehead{
  \colhead{PSR} &
  \colhead{$P$} &
  \colhead{DM} &
  \colhead{$S_{1400}$} &
  \colhead{Discovery} \\
  \colhead{} &
  \colhead{(s)} &
  \colhead{(pc\,cm$^{-3}$)} &
  \colhead{($\mu$Jy)} &
  \colhead{reference}
}

\startdata
J0040$-$7337          & 0.060          & 101.79         & 8.6          & (1) \\
J0048$-$7317          & 0.079          & 292.42         & 22           & (1) \\
J0040$-$7335          & 0.145          & 198.27         & 16           & (1) \\
J0051$-$7204          & 0.191          & 159.30         & 39           & (2) \\
J0054$-$7228          & 0.291          & 92.40          & 15           & (1) \\
J0105$-$7208          & 0.307          & 120.40         & 16           & (1) \\
J0113$-$7220          & 0.326          & 125.49         & 189          & (3) \\
J0131$-$7310          & 0.348          & 205.22         & 59           & (4) \\
J0044$-$7314          & 0.397          & 78.60          & 21           & (1) \\
J0040$-$7326          & 0.399          & 85.70          & 12           & (1) \\
J0045$-$7042          & 0.632          & 71.30          & 64           & (4) \\
J0111$-$7131          & 0.689          & 79.40          & 44           & (4) \\
J0045$-$7319          & 0.926          & 104.60         & 250          & (5) \\
J0043$-$7319          & 0.937          & 120.80         & 47           & (2) \\
\textbf{J0103$-$72}  & \textbf{1.327} & \textbf{272(13)} & \textbf{39} & \textbf{(6)} \\
\hline
J0058$-$7218          & 0.022          & \nodata        & \nodata      & (7) \\
J0100$-$7211          & 8.020          & \nodata        & \nodata      & (8) \\
\enddata

\tablecomments{
Discovery references are:
(1) \citet{carli2024};
(2) \citet{titus2019};
(3) \citet{crawford2001};
(4) \citet{manchester2006};
(5) \citet{mcconnell1991};
(6) this work;
(7) \citet{maitra2021}; and
(8) \citet{lamb2002}.
The two entries below the line are rotation-powered pulsars detected only at
X-ray wavelengths, with no radio detection and hence no measured DM.
PSR~J0103$-$72 is therefore the 15th radio pulsar known in the SMC.
}

\end{deluxetable}

\section{Observations}

\begin{figure*}[ht!]
\includegraphics[width=\textwidth]{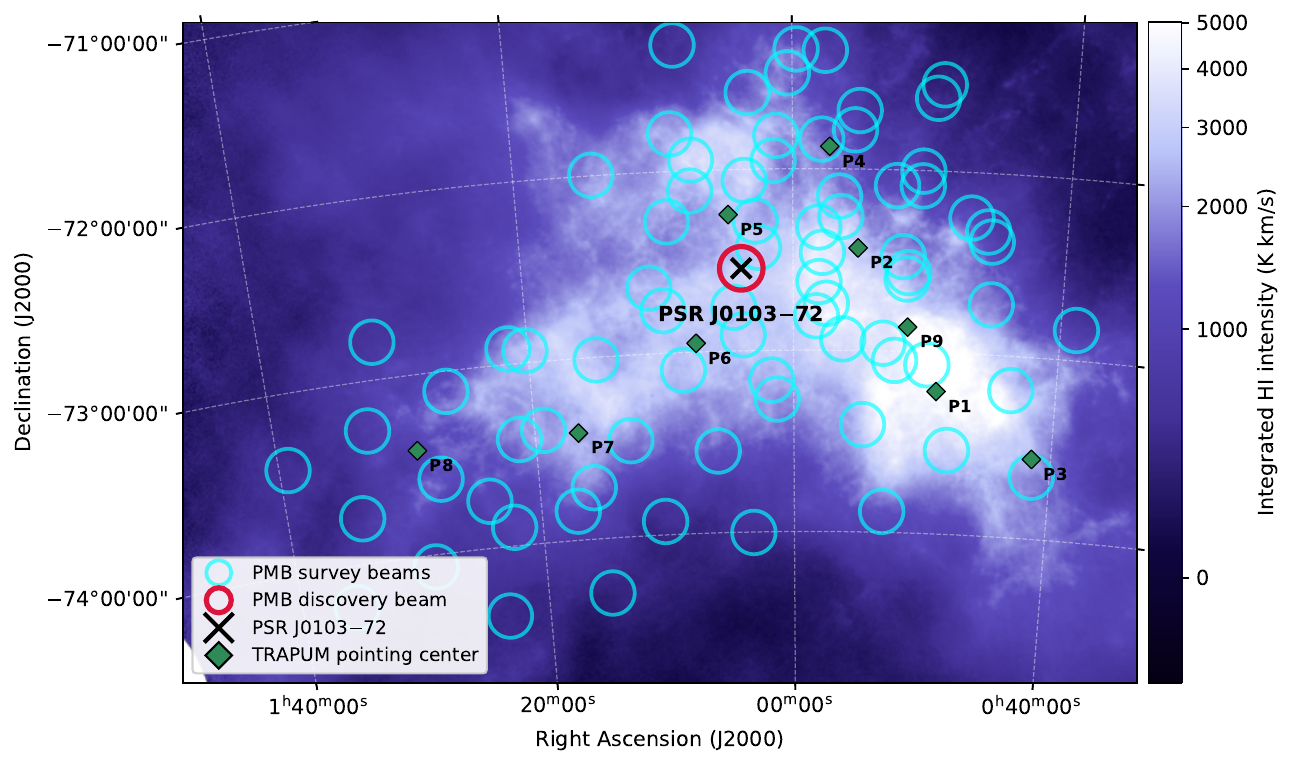}
\caption{The cyan circles show the Parkes multibeam (PMB) coverage of the 14 pointings targeted at known supernova remnants (SNRs) and pulsar wind nebulae (PWNe) in the SMC by \citet{titus2019}. The red circle marks the PMB beam in which PSR J0103$-$72 was discovered, while the black cross marks its nominal position, which is uncertain by approximately $7$~arcmin. The green diamonds, labelled P1--P8, mark the centres of the TRAPUM survey pointings from \citet{carli2024}. PSR J0103$-$72 did not fall within the coherent-beam coverage of these pointings. The background shows the integrated H\,{\sc i} intensity from the
primary-beam-corrected GASKAP-H\,{\sc i} ASKAP--Parkes SMC moment-zero
map \citep{pingel2021data, pingel2022}.}
\label{fig:pointings_trapum}
\end{figure*}

The archival data reprocessed in this work were obtained from the ATNF Data Access Portal \citep[DAP;][]{toomey2025} under project ID P944. These observations were conducted using the 64-m Parkes radio telescope (Murriyang) as part of the targeted search for young radio pulsars in the SMC by \citet{titus2019} between 2017 August 25 and 2018 March 20. Data were recorded with the Berkeley--Parkes--Swinburne Recorder (BPSR; \citealt{keith2010}) at a centre frequency of 1382\,MHz, with 400\,MHz of bandwidth divided into 1024 channels and a sampling interval of 64\,$\mu$s. In this survey, using the 13-beam 21-cm multibeam receiver \citep{staveley-smith1996}, 14 pointings were directed at known supernova remnants (SNRs) and pulsar wind nebulae (PWNe), with angular extents less than the $\sim$14\,arcmin half-power beam width of the receiver, as shown in Fig.\ref{fig:pointings_trapum}. Integration times were on average $\sim$15000\,s per pointing. All targets were observed twice, with the exception of the MCSNR~J0127$-$7332 and SNR~C2 pointings which were observed three times. The first repeated epoch were separated by $\sim$20\,days on average, allowing the confirmation of candidates between multiple observations. The original analysis of these data led to the discovery of two radio pulsars, PSR~J0043$-$7319 and PSR~J0051$-$7204 (J0043$-$73 and J0052$-$72 at discovery, and subsequently localized by \citealt{carli2024}).

\section{Data reprocessing}

\subsection{Preprocessing}

The archival data set was processed on the large computing cluster Atlas at the Albert Einstein Institute. \footnote{\url{https://www.aei.mpg.de/atlas}} Prior to the main processing, the \textsc{psrfits} format files of the survey were first converted to filterbank format and downsampled to 512 channels and a temporal resolution of 128\,$\mu$s using the \textsc{digifil} tool of \textsc{dspsr} \citep{vanstraten2011}, to reduce the computational cost posed by the long integration time. To reduce the impact of radio frequency interference (RFI) in time-domain radio data, the filterbanks were cleaned using \textsc{rfifind} in the \textsc{presto} package \citep{ransom2001}, which created a channel mask file for each beam, resulting in an average of 9.2$\%$ of the 512 channels being removed from the analysis.

\subsection{Periodicity search}

The downsampled data were searched multiple times with a pulsar search pipeline in which the FFT-based GPU-accelerated search code \textsc{peasoup}\footnote{\url{https://github.com/ewanbarr/peasoup}} \citep{peasoup} was employed using different search parameters. We first used the \textsc{peasoup} processing package to search an average of 1093 DMs per pointing from 10 to 500\,pc\,cm$^{-3}$ with a DM tolerance of 1.03. We allowed the pipeline to generate candidates with a minimum spectral S/N of 5. No restriction was placed on the period range at the search stage and the full range of periods accessible (up to 5\,s due to \textsc{peasoup}’s compiled default minimum frequency) to the Fourier search was retained. Periods beyond 5\,s in these data were searched by \citet{titus2019} with a fast folding algorithm ({FFA}; \citealt{staelin1969}), but we did not duplicate this search. We chose the lower bound of 10 $\rm pc\,cm^{-3}$ on the DM because there is no known Galactic pulsar in the direction of the SMC with a DM less than 19 $\rm pc\,cm^{-3}$. Apart from this, candidates at lower DMs are unlikely to be associated with the SMC and are also increasingly dominated by RFI. The upper limit of 500 $\rm pc\,cm^{-3}$ is well above the known SMC pulsar population and is reasonable for pulsars with larger DMs due to the SMC interstellar medium or the local environments of individual sources. We searched the data in two steps. In the first step, only periodicity search for isolated pulsars was performed without any acceleration search for binary pulsars. After the FFT search, the harmonically related Fourier amplitudes were summed up to 32 harmonics to make a peak detection and a ranked candidate lists  with a minimum spectral S/N of 5 were generated.

In the second step, an acceleration search was performed by resampling the observations over constant line-of-sight accelerations of up to $\pm10~\mathrm{m\,s^{-2}}$ with 626 trial accelerations per trial DM. The constant-acceleration approximation is expected to remain valid when $T_{\mathrm{obs}}\lesssim0.1P_{\mathrm b}$\citep{johnston1991,ransom2003} where $T_{\mathrm{obs}}$ is the integration time and $P_{\mathrm b}$ is the binary orbital period. For the mean full integration time of $T_{\mathrm{obs}}=15000~\mathrm{s}$, this corresponds to orbital periods $ \sim P_{\mathrm b}\gtrsim41.7~\mathrm{h}$, covering 59.7\% of all pulsar binaries and 74.2 \% of binaries with a spin period longer than 5\,ms (spin period range covered in this reprocessing) according to version 2.8.1 of the ATNF Pulsar Catalogue. For circular, edge-on orbits, the acceleration search is sensitive to companion masses of up to $\sim4.3\,M_\odot$ within the searched acceleration range according to Kepler's third law. This acceleration range is sensitive to the pulsar binary systems with massive stellar companions. For instance, the SMC hosts a large population of high-mass X-ray binaries, particularly Be/X-ray systems \citep{haberl2016}, which generally contain neutron stars in comparatively wide orbits around massive stellar companions \citep{ziolkowski2002, reig2011}. Currently, the only known binary pulsar in the SMC is PSR~J0045$-$7319 \citep{mcconnell1991}, which orbits a massive star of $\sim8.8 \, M_\odot$  with $P_{\mathrm b}=51.17$~d and eccentricity, $e\simeq0.81$ \citep{kaspi1994}. Using its measured orbital parameters, the maximum projected line-of-sight acceleration is only $\simeq2.7~\mathrm{m\,s^{-2}}$.

In order to preserve the sensitivity from the full integration time, we did not conduct any segmented acceleration search. Therefore, our acceleration search does not probe very compact binaries, but the chosen range provides good sensitivity to the wider massive-star binaries that are expected to be present in the SMC.  To limit the number of candidates to a computationally realistic number, candidates with a minimum spectral S/N of 6 were generated in our acceleration search.

This analysis was restricted to FFT based periodicity and acceleration searches. No single-pulse search or independent blind FFA search was performed, with the FFA used only to validate selected FFT candidates. Applying these complementary searches may reveal additional intermittent or long-period pulsars in the data.

\subsection{Candidate sifting}
\label{sec:sifting}

In our search, the integration length of the data, combined with DM and acceleration trials while generating candidates down to a low spectral S/N threshold, produced millions of candidates. Therefore, it is computationally very demanding and impractical to fold and optimize each and every candidate, making candidate selection an essential stage  because majority of the candidates are spurious due to  random fluctuations of background noise as well as RFI \citep{lyon2016}. In order to remove such noise candidates and limit the number of candidates for folding, FFT-based search pipelines often use a uniform $\mathrm{S/N_{FFT}}$ threshold to reduce the number of candidates to be folded to a manageable number. As narrow periodic signals distribute power across various harmonics in the Fourier domain \citep{bracewell2000}, a uniform threshold might cause narrow duty-cycle pulsars to be missed, especially so for faint slow pulsars. The incoherent summing of harmonics can lower $\mathrm{S/N_{FFT}}$ even further \citep{ransomeikenberry2002}. As described by \citet{sengar2023}, on average the folded S/Ns for faint narrow duty-cycle pulsars can be 1.5--2.0 times or even higher than their spectral S/Ns. In view of such challenges, an adaptive $\mathrm{S/N_{FFT}}$ threshold for candidate sifting based on spin period and the number of harmonic sums ($n_{h}$) corresponding to each candidate was used as follows:

For the initial zero-acceleration search, a candidate was selected if it satisfied any one of the following criteria:

\begin{equation}
\label{eq:candidate_selection}
\begin{array}{ccc}
\hline
\text{Period range} & \text{Harmonics} & \mathrm{S/N}_{\mathrm{FFT,min}} \\
\hline
\text{all }P                  & n_h=0           & 9.5 \\
\text{all }P                  & n_h\geq1         & 7.5 \\
30\leq P\leq100~\mathrm{ms}  & 1\leq n_h\leq4  & 7.0 \\
P>100~\mathrm{ms}            & n_h\geq1         & 7.0 \\
50\leq P\leq200~\mathrm{ms}  & n_h\geq3         & 5.0 \\
200<P<300~\mathrm{ms}        & n_h\geq3         & 6.0 \\
300\leq P<500~\mathrm{ms}    & n_h\geq3         & 5.5 \\
P\geq500~\mathrm{ms}         & n_h\geq3         & 5.0 \\
\hline
\end{array}
\end{equation}

The same criteria were used for the acceleration search, except that the low-threshold criterion was omitted for candidates with $50\leq P\leq200~\mathrm{ms}$, $n_h\geq3$, and $\mathrm{S/N}_{\mathrm{FFT}}\geq5.0$. Applying this criterion to every acceleration trial would significantly increase the number of candidates selected for folding and was therefore computationally prohibitive.

In the zero acceleration periodicity search, out of $\sim$13\,550 candidates per beam from the \textsc{peasoup} pipeline candidate list, on average 235 candidates per beam successfully passed through the candidate selection process for folding, leading to a total of $\sim$42770 candidates for folding from all targeted observations. 

However, in the acceleration search, due to the large number of acceleration trials, $\sim$120500 candidates per beam were generated in the candidate lists, and even after applying the sifting conditions, an average of $\sim$52571 candidates per beam was left, leading to a total of $\sim$ 9567900 candidates for folding from all targeted observations. Upon further inspection, we found that candidates with periods less than 5\,ms had severe RFI dominance, with the majority of the brightest spurious signals having periods less than 1\,ms, as well as the degraded sensitivity to millisecond pulsars arising from the $\sim$1\,ms intra-channel smearing at the top of the searched DM range combined with the 128\,$\mu$s sampling \citep{lorimer2004}.

Apart from this we also found that this large number of candidates was partly due to persistent periodic RFI that remained present for 10--20 minutes in several observations. A single long-duration RFI signal can be detected repeatedly across many DM and acceleration trials, and produce a large number of related candidates. Therefore, to further reduce the number of candidates in the acceleration search, we identified RFI stripes in the period--DM plane. Terrestrial periodic interference typically remains at an approximately fixed Fourier frequency\citep{maan2021} while appearing in many DM trials \citep{eatough2009}, whereas the signal of a pulsar generally peaks within a more restricted DM range \citep{lorimer2004}. Candidates were therefore grouped into frequency stripes with a width corresponding to twice the Fourier-bin resolution, $\Delta f_{\mathrm{stripe}}=2/T_{\mathrm{obs}}$.  Since $f = \frac{1}{P}$, a fixed width in frequency corresponds to a period width $\Delta P\simeq P^2\Delta f$ for $\Delta f\ll f$.
Hence the approximate stripe width in period was
\begin{equation}
\Delta P_{\mathrm{stripe}}
\simeq \frac{2P^2}{T_{\mathrm{obs}}},
\end{equation}

The stripe is therefore narrow in period for short-period candidates and broad for long-period ones.

A stripe was classified as an RFI stripe when it contained at least 50 candidates spanning a DM range of $70~\mathrm{pc\,cm^{-3}}$. Candidates belonging to stripes satisfying these criteria were rejected.  As a result, on average 6\,305 candidates per beam were selected for folding out of $\sim$120500 candidates per beam from the \textsc{peasoup} pipeline candidate list.

\subsection{Folding and inspection}

Sifted candidates underwent folding using \textsc{dspsr}
\citep{vanstraten2011} to create folded archive files which were cleaned using \textsc{clfd}\footnote{\url{https://github.com/v-morello/clfd}} \citep{morello2019, morello2023}. The folded diagnostic plots were generated using the \textsc{pdmp} module of \textsc{psrchive} \citep{hotan2004}. For the initial zero acceleration search, 4055 candidates with folded S/N greater than 8.0 were selected for visual inspection, as candidates with lower folded S/N would be very difficult to confirm given the long duration of observations. For the acceleration search, given the large number of acceleration trials, a higher folded S/N threshold of 9.0 was applied for visual inspection to further remove spurious candidates. This resulted in a final number of 11349 candidates for visual inspection.

The known SMC pulsars have DMs spanning $71$--$292~\mathrm{pc\,cm^{-3}}$, indicating that radio pulsars in the SMC are significantly dispersed. A compelling SMC pulsar candidate should therefore
be weaker when refolded at $\mathrm{DM}=0$, whereas terrestrial RFI generally lacks this dispersion signature of radio pulsars \citep{eatough2009}. Promising candidates were therefore refolded at zero DM, and the resulting diagnostic plots were visually compared with those obtained at the optimised DM to identify possible RFI. Where additional observations of the same pointing were available, they were also folded at the candidate period and DM for independent confirmation.

\subsection{Sensitivity}
\label{sec:sensitivity}
In order to compare the radio flux density sensitivity limits of the SMC surveys, we used the radiometer equation applied to pulsar observations \citep[p.~265]{lorimer2004}:

\begin{equation}
S_{\mathrm{min}} =
  \frac{(\mathrm{S/N})_{\mathrm{min}}\,(T_{\mathrm{sys}} + T_{\mathrm{sky}})\,\beta}
       {G\,\sqrt{n_{\mathrm{pol}}\,t_{\mathrm{obs}}\,\Delta\nu}}
  \sqrt{\frac{\delta_{\mathrm{eff}}}{1 - \delta_{\mathrm{eff}}}},
\label{eq:radiometer}
\end{equation}

\noindent
 with parameters defined as shown in Table \ref{tab:sensitivity} where $\delta_{\mathrm{eff}} = W_{\mathrm{eff}}/P$ and the effective pulse width accounts for broadening by the sampling time, intra-channel dispersion smearing, the dedispersion step and scattering:

\begin{equation}
W_{\mathrm{eff}} = \sqrt{W_{\mathrm{int}}^{2} + t_{\mathrm{samp}}^{2}
  + t_{\mathrm{chan}}^{2} + t_{\mathrm{DM}}^{2} + t_{\mathrm{scatt}}^{2}}.
\label{eq:weff}
\end{equation}

\noindent

\begin{figure}[ht!]
\includegraphics[width=\columnwidth]{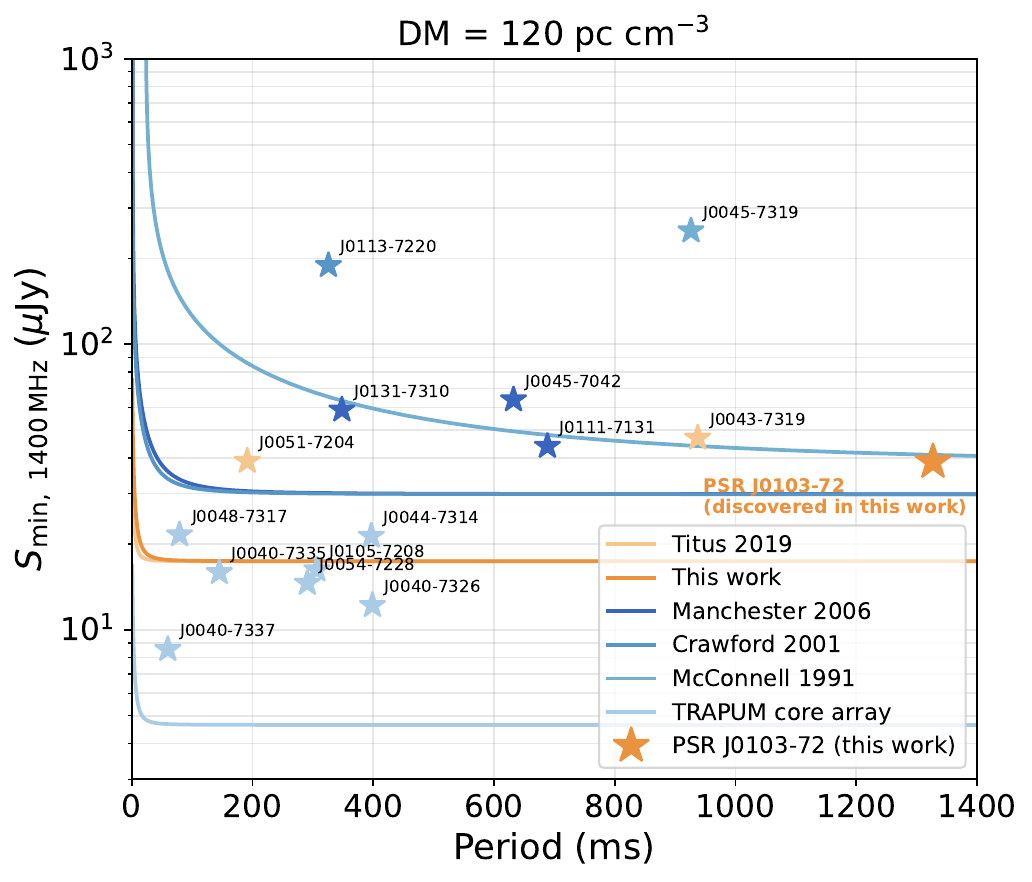}
\caption{The minimum flux density sensitivity curves and the pulsar
discovery flux densities (modified by beam losses) for the pulsar surveys of the SMC. All 15 known SMC radio pulsars in Table \ref{tab:smc} are plotted. The calculations used to produce this figure and interpretations are detailed in Section \ref{sec:sensitivity}.}
\label{fig:sensitivity}
\end{figure}

In Fig.~\ref{fig:sensitivity} we show the flux density sensitivity of all the SMC pulsar surveys calculated with Equation~\ref{eq:radiometer} for the median DM ($120 \rm pc \, \rm cm^{-3}$) of the known SMC radio pulsars. The parameters for these surveys are taken from their respective publications, which we list in Table~\ref{tab:sensitivity}.

For all surveys, we assume a minimum folded S/N of 9 to be confidently
identified by eye inspection, given that no known SMC pulsar was discovered with a folded S/N lower than 9. We assume the gain of Murriyang to be 0.735\,K\,Jy$^{-1}$, the boresight value \citep{manchester2001}, in all cases. The digitisation correction factors are taken from \citet{kouwenhoven2001}: 1.00 for 8-bit digitisation, 1.25 for 1-bit and 1.06 for 2-bit. The sky temperature in the direction of the SMC at each survey's centre frequency is derived using the \citet{zheng2017} model via \textsc{pygsm} \citep{price2016}, with the contribution from the cosmic microwave background taken to be 2.72548\,K \citep{fixsen2009}. We do not apply an additional efficiency factor to convert between spectral and folded S/N, since the thresholds adopted here are quoted as folded S/N throughout.

All survey flux density sensitivities are scaled to 1400\,MHz assuming a pulsar spectral index of $-1.60$ \citep{jankowski2018}. Among these, the
\citet{mcconnell1991} survey had a center frequency of 610\,MHz, outside the L band. The spectral index conversion is therefore more uncertain for this survey, as the range of pulsar spectral indices may change the resulting flux density upper limit. The other surveys compared were all conducted at L band, close to a center frequency of 1400\,MHz, and the conversion is correspondingly more accurate. We use the median DM of the SMC radio pulsars listed in Table~\ref{tab:smc}, 120\,pc\,cm$^{-3}$, and assume an intrinsic pulsar duty cycle of 2.5 per cent, 

We also account for pulse broadening due to scattering from multipath
propagation through the interstellar medium, using Equation~7 and the fitted constants of \citet{bhat2004}. It should be noted, however, that this relation was calibrated on pulse-broadening measurements of several hundred Galactic pulsars along typical Galactic sightlines, making it only a physically motivated estimate when extrapolated to the SMC. Given that the Galactic foreground contributes only 30\,pc\,cm$^{-3}$ along this sightline \citep{yao2017}, most of the observed DM arises within the SMC itself rather than in the diffuse Galactic ISM on which the fit was built. At the median SMC DM of 120\,pc\,cm$^{-3}$ and $\nu = 1382$\,MHz, the relation gives $t_{\mathrm{scatt}} \approx 0.009$\,ms, more than an order of magnitude smaller than the sampling and intra-channel smearing terms. Furthermore, the thin-screen scattering delay carries a geometric factor $d_{\mathrm{s}}(D - d_{\mathrm{s}})/D$, where $d_{\mathrm{s}}$ is the observer--screen distance and $D$ the observer--pulsar distance, which tends to zero as the screen approaches the pulsar. Material internal to the SMC therefore contributes considerably less scattering than the same DM would in a mid-path Galactic screen \citep[eq.~C2]{cordes1998}, so the true $t_{\mathrm{scatt}}$ is smaller still and our treatment is conservative. As a result of these effects, $\delta_{\mathrm{eff}}$ departs from the assumed intrinsic value at short periods.

Temperature contributions from the ground and the atmosphere, scintillation, the effective bandwidth after RFI masking, and harmonic summing are not taken into account, owing to the limited availability of these details across surveys.

\begin{deluxetable*}{lcccccc}
\tablecaption{Parameters of the SMC pulsar surveys used to calculate the flux
density sensitivity limits shown in Fig.~\ref{fig:sensitivity}, following
Equation~\ref{eq:radiometer}. Candidates in this work were inspected down to a
folded S/N of 8, but we adopt 9 throughout as the level at which a signal is
confidently identified by eye.\label{tab:sensitivity}}

\tablehead{
\colhead{Parameter} &
\colhead{(1)} &
\colhead{(2)} &
\colhead{(3)} &
\colhead{(4)} &
\colhead{(5)} &
\colhead{(6)}
}

\startdata
Minimum folded S/N, $(\mathrm{S/N})_{\mathrm{min}}$
    & 9.0   & 9.0   & 9.0  & 9.0  & 9.0   & 9.0   \\
Centre frequency $\nu$ (MHz)
    & 1382  & 1382  & 1374 & 1374 & 610   & 1284  \\
Bandwidth $\Delta\nu$ (MHz)
    & 400   & 400   & 288  & 288  & 120   & 856   \\
Number of frequency channels
    & 512   & 1024  & 96   & 96   & 24    & 2048  \\
System temperature $T_{\rm sys}$ (K)
    & 23.0  & 23.0  & 21.0 & 21.0 & 44.0  & 18.0  \\
Sky temperature $T_{\rm sky}$ (K)
    & 6.58  & 6.58  & 6.61 & 6.61 & 18.78 & 7.01  \\
Sampling time $t_{\rm samp}$ ($\mu$s)
    & 128   & 64    & 1000 & 250  & 5000  & 153   \\
Digitisation correction $\beta$
    & 1.06  & 1.06  & 1.25 & 1.25 & 1.25  & 1.00  \\
Integration time $t_{\rm obs}$ (s)
    & 15000 & 15000 & 8400 & 8400 & 5000  & 7200  \\
Gain $G$ (K\,Jy$^{-1}$)
    & 0.735 & 0.735 & 0.735 & 0.735 & 0.735 & 1.925 \\
Assumed duty cycle $\delta$ (per cent)
    & 2.5   & 2.5   & 2.5  & 2.5  & 2.5   & 2.5   \\
Number of polarizations recorded $n_{pol}$
    & 2     & 2     & 2    & 2    & 2     & 2     \\
\enddata

\tablecomments{
Survey columns are:
(1) this work;
(2) the targeted SMC survey of \citet{titus2019};
(3) the survey of \citet{manchester2006};
(4) the survey of \citet{crawford2001};
(5) the survey of \citet{mcconnell1991}; and
(6) the TRAPUM core survey described by \citet{carli2024}.
}

\end{deluxetable*}

\section{Results}

\subsection{New discovery}

\begin{figure*}[ht!]
\centering

\includegraphics[width=1.0\textwidth]{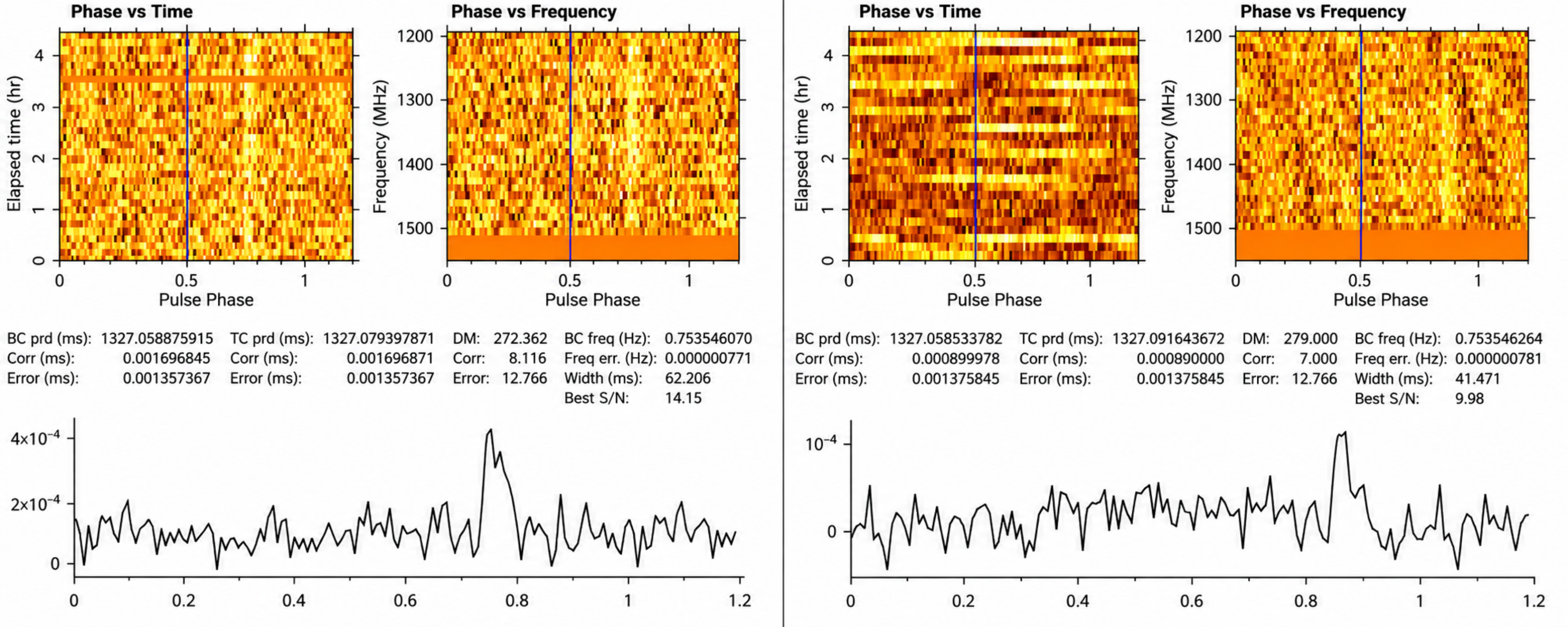}
\caption{Discovery and confirmation folds of PSR~J0103$-$72, produced with \textsc{pdmp}. \textit{Left:} the discovery observation of 28 August 2017 (beam 07), folded after manual excision of an RFI stripe, giving a folded S/N of 14.15. \textit{Right:} the confirmation observation of 12 September 2017 (beam 06), folded at the discovery period and DM, giving a folded S/N of 9.98; this observation is more heavily affected by RFI. In each panel, the upper plots show pulse phase against time and against observing frequency, and the lower plot shows the integrated pulse profile.}
\label{fig:discovery}
\end{figure*}

From the inspection of non-accelerated candidates we found a very compelling candidate. It was first detected in the 2017 August 28 pointing with a \textsc{pdmp} folded S/N of 14.2 after manually removing an RFI stripe. The pulsar was detected in one of the inner ring beams of the multibeam receiver, with beam coordinates RA\,$=$\,01:03:48.84 and Dec\,$=$\,$-$72:32:44.0 (beam 07) as shown in Fig. \ref{fig:pointings_trapum}. Hereafter, we will refer to this pulsar as PSR~J0103$-$72. It was detected again in a second epoch with a folded S/N of 9.98 in the 2017 September 12 pointing, with beam coordinates RA\,$=$\,01:03:48.58 and Dec\,$=$\,$-$72:32:44.0 (beam 06). The S/N difference between the two detections is mainly due to the heavy RFI contamination in the second
detection. However, other factors, such as scintillation and beam position, may also contribute. Despite this contamination, the pulsar is clearly visible in the confirmation diagnostic plot (see Fig. \ref{fig:discovery}). We cross-checked the nearby pulsars using \texttt{pulsar survey scraper}\footnote{\url{https://pulsar.cgca-hub.org/}} \citep{pss} and found no match. The nearest pulsar to PSR~J0103$-$72 is PSR J0050$-$7250 which is 1.07 deg away.

PSR~J0103$-$72 has a DM of 272\,pc\,cm$^{-3}$, which is consistent with the known SMC pulsar DM range (71.3--292.4\,pc\,cm$^{-3}$; Table~\ref{tab:smc}) and makes it the second highest known in the SMC after PSR~J0048$-$7317, whose DM is 292.4\,pc\,cm$^{-3}$ \citep{carli2024}. This pulsar has a spin period of
1.327\,s, making it the slowest radio pulsar known in the SMC.

Using the pulse profile from the first epoch observation, we derived its flux density from Equation \ref{eq:radiometer}, $S_{1400} \approx 39\,\mu \, \rm Jy$, using the parameters listed for this work in Table~\ref{tab:sensitivity}. Adopting an SMC distance of 62.44\,kpc \citep{graczyk2020}, the corresponding pseudo-luminosity is $L_{1400} = S_{1400}\,d^{2} \approx 150 $\,mJy\,kpc$^{2}$. This value is at the median of the pseudo-luminosity of all SMC radio pulsars, with $L_{1400}$ value of 7 radio pulsars discovered in \citet{carli2024} calculated with each discovery's own S/N and duty cycle calculated from W50 and P in ATNF catalogue, gain scaled linearly with antenna count, rescaled to 1400 MHz, and $L_{1400}$ value of older discoveries taken from ATNF.

The position of the pulsar within the $\sim$14\,arcmin beam is unknown, hence the flux calculated is a lower limit estimate without beam attenuation correction taken into account. If the pulsar lies at the half-power point, $S_{1400} \approx 78\,\mu \, \rm Jy$.

The confirmation epoch can also be used to calculate the flux density $S_{1400} \approx 22.13\,\mu$Jy corresponding to a  pseudo luminosity of $L_{1400} = S_{1400}\,d^{2} \approx 86.29$\,mJy\,kpc$^{2}$ in the same manner. This lower value is partially because no pulsed emission is detected between 1200 and 1300\,MHz during this epoch, most likely due to interstellar scintillation. However, we note that this calculation is not as reliable, given that the folded S/N is low due to heavy RFI contamination. No known SNR or PWN is coincident with the beam centre. The nearest, MCSNR~J0103$-$7247, lies 14.9 arcmin away \citep{cotton2024}.

PSR~J0103$-$72 was detected in the zero-acceleration search and in the acceleration search, the highest S/N was obtained at an acceleration of $-0.07~\mathrm{m\,s^{-2}}$, which is consistent with zero at the resolution of the search. Moreover, neither the discovery nor the confirmation observation plot shows an apparent phase drift attributable to acceleration. Therefore, PSR~J0103$-$72 is most likely an isolated pulsar.

\begin{figure*}[ht!]
    \centering
    \includegraphics[width=\textwidth]{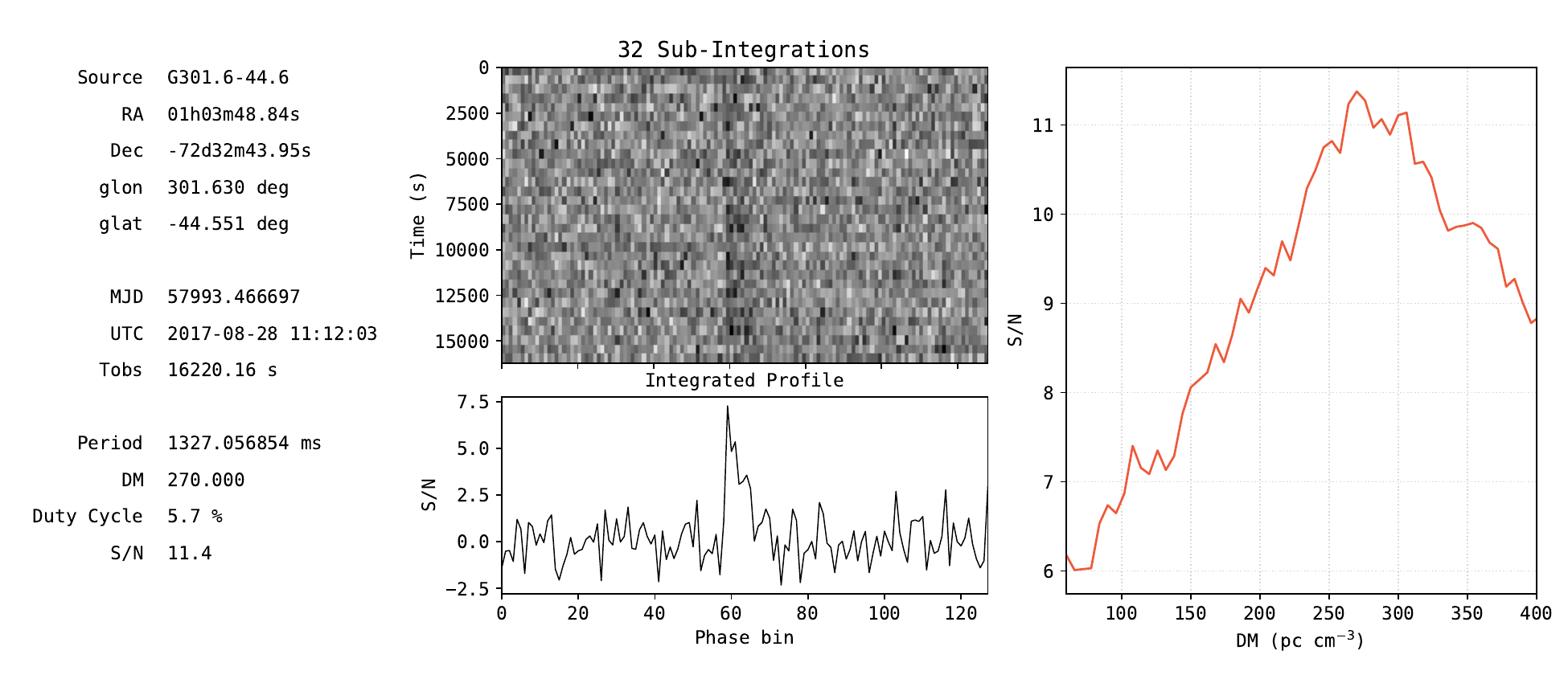}
    \caption{Redetection plot of PSR J0103$-$72 in the discovery observation using \texttt{RIPTIDE}. The DM search range spans from 0 to 400 \,pc\,cm$^{-3}$ with DM steps of 2\,pc\,cm$^{-3}$ as described by \citet{titus2019}.}
    \label{fig:ffa}
\end{figure*}
\subsection{Detection of known pulsars}

Our FFT-based pipeline not only discovered PSR~J0103$-$72 but also detected three known pulsars, including both PSR~J0043$-$7319 and PSR~J0051$-$7204, discovered in the original processing of this survey by \citet{titus2019}. PSR~J0043$-$7319 was originally discovered only using FFA, but was identified as a candidate pulsar in our search when the number of summed harmonics was raised to 16. We detected its second harmonic with an $\mathrm{S/N_{fold}}$ of 11.22. The fundamental can be recovered at an $\mathrm{S/N_{fold}}$ of 13.37 by folding the raw data at the DM obtained from the harmonic candidates over a range of harmonically related periods. The folded profiles of both pulsars are clearly visible in the diagnostic plots. We also detected PSR~J0051$-$7204 with an $\mathrm{S/N_{fold}}$ of 10.01. PSR~J0057$-$7201, a foreground Galactic pulsar originally discovered in a survey by \citet{crawford2001}, was also detected with an $\mathrm{S/N_{fold}}$ of 41.31.

\section{Discussion}

\subsection{The previous non-detection of PSR~J0103$-$72}
\label{sec:previous}

As shown in Fig.~\ref{fig:sensitivity}, with a flux density $S_{1400} \approx 39\,\mu$Jy, PSR~J0103$-$72 lies well above the sensitivity limit of the original survey. Sensitivity is therefore not the reason for its previous non-detection.

An FFA search was also used in \citet{titus2019} to search for pulsars with periods from 200\,ms to 360\,s, and DMs from 0 to 400\,pc\,cm$^{-3}$ with DM steps of 2\,pc\,cm$^{-3}$. We attempted to reproduce this by conducting an FFA search using \texttt{RIPTIDE} \citep{morello2020} with the same period range, DM range and DM steps adopted by \citet{titus2019}. This search was able to recover the pulsar with  FFA S/N of 11.4  as shown in Fig.~\ref{fig:ffa} and folded $\mathrm{S/N = 14.0}$ after optimization by \textsc{pdmp}. For comparison, PSR~J0043$-$7319, which was discovered by FFA in the same survey, has a search-stage S/N of 13.5. PSR~J0103$-$72 is therefore fainter than the only FFA discovery reported from these data, although \citet{titus2019} do not state an explicit FFA threshold.

The initial diagnostic plot that allowed the discovery of PSR~J0103$-$72 contained a bright horizontal RFI stripe in time. The frequency was later manually zapped in Fig.~\ref{fig:discovery}. The second observation for confirmation was heavily contaminated by RFI. The heavy RFI contamination could have contributed to its non detection in previous search. The processing by \citet{titus2019} adopted a significance of 4.0$\sigma$ for candidates to be folded, which yielded $\sim$150000 FFT candidates and a further 40\,000 FFA candidates for visual inspection. The sheer volume of candidates requiring inspection by eye may have contributed to the non-detection. However, a uniform threshold to reduce the candidate volume on $\mathrm{S/N_{FFT}}$ risks discarding narrow duty-cycle pulsars.

A direct comparison between the thresholds used in the previous FFT based search using \texttt{PRESTO} and the spectral S/N measured in our \texttt{PEASOUP} search is not possible because \texttt{PRESTO} and \texttt{PEASOUP} measure and normalize their detection statistics differently. \texttt{PRESTO} uses Gaussian significance of the harmonic summed power spectrum \citep{ransomeikenberry2002} and \texttt{PEASOUP} uses S/N measured from the normalized harmonic summed power spectrum \citep{sengar2023}. In our search, PSR~J0103$-$72 was found with a spectral S/N of 6.7 when 32 harmonics were summed, so it is likely that if harmonic summing up to 16 is used, the pulsars would have an even lower spectral S/N. The earlier \texttt{PRESTO} search summed only up to 16 harmonics. Since the pulsar was detected most strongly after summing up to 32 harmonics, its Gaussian significance may have been lower than $4\sigma$ in the earlier search, and the pulsar would not have survived a uniform cut placed high enough to bring the candidate list to a manageable size.  Under our candidate selection criteria, the pulsar was found  through one of the long-period selection criteria which requires $P \geq 200$\,ms and $n_{h} \geq 3$. This threshold significantly reduced the number of candidates inspected in this work to 4055 for zero acceleration search. Given that \citet{titus2019} did not conduct any acceleration search, the additional 11349 candidates for visual inspection in acceleration search should not be used for a fair comparison.

\begin{figure*}[t!]
    \centering
    \includegraphics[width=0.49\textwidth]{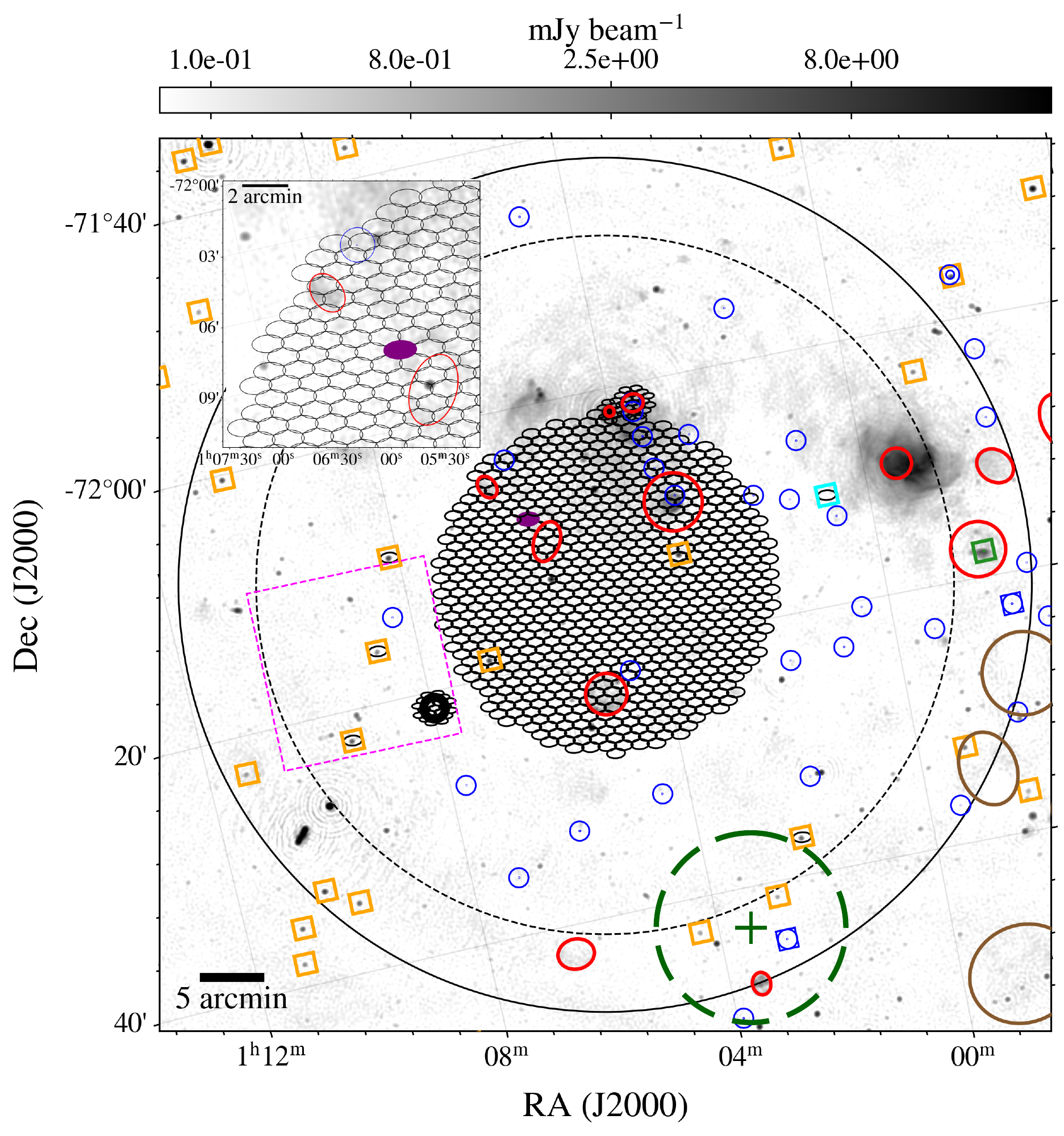}%
    \hfill%
    \includegraphics[width=0.49\textwidth]{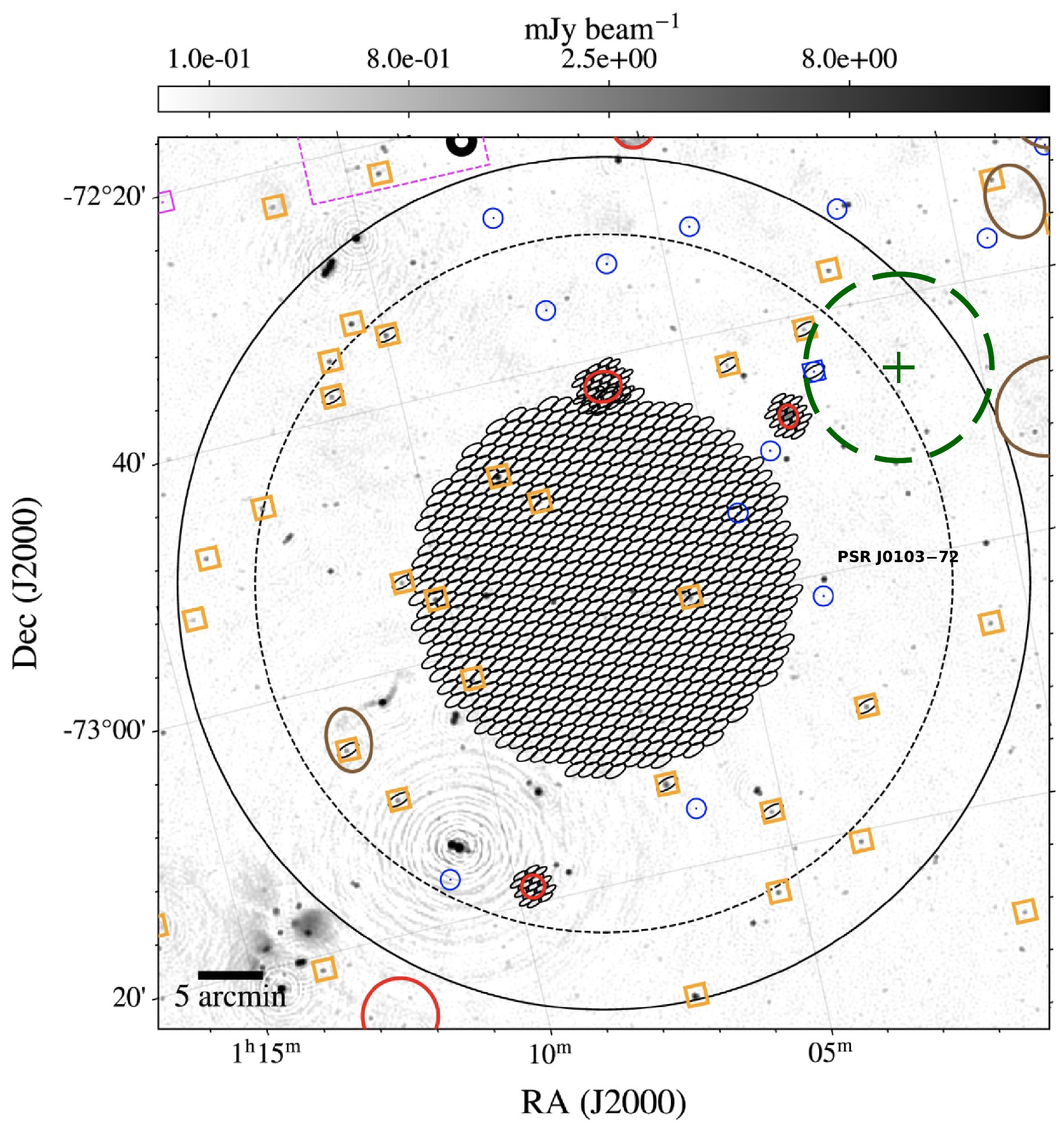}

    \caption{The Parkes multibeam discovery beam of PSR J0103$-$72 overlaid on the beam layout maps of the two nearest TRAPUM SMC survey pointings as shown in Fig.~\ref{fig:pointings_trapum}. The dashed dark green circle shows the $14$~arcmin FWHM (half-power) contour of the Parkes discovery beam with the beam center marked with a cross. The large solid and dashed black circles show the $1.1^{\circ}$ and $0.9^{\circ}$ FWHM contours of the MeerKAT incoherent (primary) beam at the center and top frequency of L band, respectively \citep{asad2021}. The small black ellipses represent the MeerKAT core array coherent beams. \textit{Left:} SMCPOINTING5. \textit{Right:} the first pass of SMCPOINTING6, the second pass had a similar beam layout and is not shown. In both observations, the discovery beam of PSR J0103$-$72 overlaps the MeerKAT incoherent-beam but lies outside the coherent-beam tiling. The background panels are adapted from Figs.~B7(a) (left) and B5(a) of \citet{carli2024} (right) under the Creative Commons Attribution 4.0 license; all other symbols are defined there.}
    \label{fig:trapum_beam_coverage}
\end{figure*}

The TRAPUM Small Magellanic Cloud pulsar survey \citep{carli2024} utilises the MeerKAT radio interferometer. As shown in Fig.~\ref{fig:sensitivity}, this highly sensitive survey could have detected PSR~J0103$-$72. However, the Parkes multibeam pointing in which PSR~J0103$-$72 was discovered was not covered by the coherent-beam tilings of the nearby TRAPUM survey pointings, as shown in Fig.~\ref{fig:trapum_beam_coverage}. The Parkes discovery beam nevertheless overlaps the MeerKAT incoherent-beam pointings of three TRAPUM observations, the single observation of SMCPOINTING5 and two observations of SMCPOINTING6. The TRAPUM incoherent-beam observations have a nominal 1400-MHz flux-density sensitivity limit of $S_{1400,\mathrm{min}}=37.2\pm1.8~\mu\mathrm{Jy}$ \citep{carli2024}. This is comparable to the estimated discovery-epoch flux density of PSR~J0103$-$72, $S_{1400}\approx39~\mu\mathrm{Jy}$. The pulsar would therefore have been close to the nominal detection threshold, with its actual detectability depending on its position within the incoherent beam. Given the substantial flux variability observed among SMC pulsars \citep{johnston2022}, its non-detection in these observations is not unexpected.

\subsection{Lack of association}
\label{sec:association}
 
The nearest known remnant is MCSNR~J0103$-$7247 \citep{maggi2019, cotton2024}, 14.9\,arcmin away from PSR~J0103$-$72. Given MCSNR~J0103$-$7247's age of 5.1\,kyr \citep{leahy2022}, an association would require a projected velocity of $\sim$5$\times$10$^{4}$\,km\,s$^{-1}$, assuming PSR~J0103$-$72 is located at the beam centre or $\sim$2.8$\times$10$^{4}$\,km\,s$^{-1}$, under the most favourable assumption that PSR~J0103$-$72 is at the boundary of the $\sim$14\,arcmin half-power beam width of the Parkes BPSR receiver. Both are beyond the known distribution of the birth velocities of radio pulsars \citep{lyne1994, verbunt2017}. Hence, it is unlikely for PSR~J0103$-$72 to be associated with any known SNR or PWN.

\section{Conclusion}

We have reprocessed the targeted SMC observations of \citet{titus2019} using improved and sensitive search techniques and discovered PSR J0103$-$72. The pulsar was independently detected in two observations and has a spin period of 1.327~s and a dispersion measure of $272\pm13~{\rm pc \, cm^{-3}}$. Its discovery increases the known population of radio pulsars in the SMC from 14 to 15. PSR J0103$-$72 is the slowest known radio pulsar in the SMC, which extends the upper end of the observed spin period distribution in the SMC from 0.937~s to 1.327~s (by 42\%). This discovery shows that the detectable SMC population is not restricted to young, relatively faster pulsars, including those associated with supernova remnants and pulsar-wind nebulae, but also contains longer-period objects that are likely to have been under-represented in previous searches. From the discovery observation, the pulsar has a 1400-MHz flux density of approximately $39~\mu{\rm Jy}$ and a pseudo-luminosity of $151~{\rm mJy\,kpc^{2}}$. These are lower limits because the pulsar position is uncertain by $7$~arcmin. In addition, SMC pulsars can show substantial flux variability \citep{johnston2022}. 

After subtracting the predicted Milky Way foreground contribution of $30$--$42~\mathrm{pc\,cm^{-3}}$ \citep{cordes2002, yao2017}, the measured DM leaves approximately $230$--$242~\mathrm{pc\,cm^{-3}}$ attributable to the SMC and material local to the pulsar. The only known SMC radio pulsar with a higher DM is PSR J0048$-$7317, with $\mathrm{DM}=292.42~\mathrm{pc\,cm^{-3}}$ \citep{carli2024}. Given the $13~\mathrm{pc\,cm^{-3}}$ uncertainty for PSR J0103$-$72, the difference is only $\sim1.6\sigma$. PSR J0048$-$7317 is a young, energetic pulsar with a spin period of 79 ms and associated with a bow-shock PWN, so its high DM may include a contribution from ionized material in its young local environment, although this has not yet been established. In contrast, PSR J0103$-$72 has a substantially longer spin period and no known SNR or PWN association, making such a local environment contribution less immediately apparent.

The large DM of PSR J0103$-$72 could therefore arise from enhanced ionized gas along the sightline, a long path through the diffuse ionized medium of the SMC, or a combination of the two. As a limiting case, if the excess DM is attributed entirely to a uniform medium with the YMW16 central SMC free electron density of $0.045\pm0.017~\mathrm{cm^{-3}}$ \citep{yao2017}, the corresponding line-of-sight path length through the SMC will be $5.1\pm2.0$~kpc which is comparable to SMC depth estimates of $4.3\pm1.0$~kpc and $5.2$~kpc derived from RR Lyrae stars \citep{muraveva2017}, and would imply that the pulsar is located toward the far side of the SMC's main body. 

This interpretation is not conclusive, however, because enhanced ionized gas along the sightline could produce the same electron density and hence, the same high DM over a shorter distance. Precise localization of PSR J0103$-$72 would permit comparison of its position with H$\alpha$, H\,{\sc ii}, H\,{\sc i}, and radio-continuum maps, helping to assess whether any presence of bright H$\alpha$ emission or known H{\sc ii} structures along the projected sightline could contribute to its excess DM. At $S_{1400} \approx 39\,\mu \, \rm Jy$, the pulsar is within reach of the MeerKAT coherent tiling, which reaches $6.6 \,\mu \, \rm Jy$ in 2 h \citep{carli2024}. However, follow-up work of this pulsar, including its timing, will be presented elsewhere.

The discovery of PSR J0103$-$72 shows how faint, long-period pulsars can remain unidentified in existing survey data sets. PSR J0103$-$72 was recovered with both FFT and FFA searches, suggesting that its earlier non-detection was not simply a limitation of either search method. RFI and the large number of candidates requiring inspection may have contributed. However, the original cause cannot be established. Its detection with period and harmonic dependent FFT thresholds in noise floor candidates highlights the importance of candidate selection in discovering faint pulsars. Therefore, reanalysis of archival observations remains immensely useful.

\begin{acknowledgments}
We thank the anonymous referee for their comments and suggestions, which helped improve the quality of the paper. This research was conducted during a three-month undergraduate research training programme at the Max Planck Institute for Gravitational Physics, Hannover. W.W. and R.S. are thankful for support from the Max Planck Society. We thank Lawrence Toomey for making the archival Parkes observations available through the Data Access Portal, without which this work would not have been possible. This work made use of the large computing cluster Atlas operated by the Observational Relativity and Cosmology division of the Albert Einstein Institute. The Parkes Murriyang radio telescope is part of the Australia Telescope National Facility which is funded by the Australian Government for operation as a National Facility managed by CSIRO (the Commonwealth Scientific and Industrial Research Organisation). We acknowledge the Wiradjuri people as the Traditional Owners of the Observatory site.

\end{acknowledgments}

\bibliography{main}{}
\bibliographystyle{aasjournalv7}



\end{document}